\documentclass[aps,superscriptaddress,showpacs,nofootinbib,floatfix,epfs]{revtex4}
\pdfoutput=1
\usepackage{amsfonts,amscd,amsmath, amssymb,graphicx,color}

\newcommand{\oh}{\frac{1}{2}}
\def\g3{\mathbf{g}}
\def\G3{\boldsymbol{\Gamma}}
\def\oh{\frac{1}{2}}

\def\e{\text{e}}
\def\ets{\eta^{\text{\tiny S}}}
\def\etaa{\eta^{\text{\tiny A}}}
\begin{document}
\title{More On Nonrelativistic Diffeomorphism Invariance}
\author{Oleg Andreev}
\affiliation{L.D. Landau Institute for Theoretical Physics, Kosygina 2, 119334 Moscow, Russia }
\affiliation{Arnold Sommerfeld Center for Theoretical Physics, LMU-M\"unchen, Theresienstrasse 37, 80333 M\"unchen, Germany}
\begin{abstract} 
Certain aspects of nonrelativistic diffeomorphisms in $2+1$ dimensions are investigated. These include a nonrelativistic limit of some relativistic 
actions in $3$ dimensions, the Seiberg-Witten map, a modification of the viscosity tensor in particular due to a non-uniform magnetic field, a redefinition of background fields, and $1/R$ terms on Riemann surfaces of constant curvature. 
 
\end{abstract}
\preprint{LMU-ASC 53/14}
%\date{\today}
\maketitle

\vspace{-7.5cm}
\begin{flushright}
LMU-ASC 53/14
\end{flushright}
\vspace{6cm}

\vspace{1cm}
%____________________________________________________ section 1
\section{Introduction}
\label{intro}
\renewcommand{\theequation}{1.\arabic{equation}}
\setcounter{equation}{0}

Effective field theory is more than a convenience. This is an appropriate description of the important physics at a certain corner of the parameter 
space of the world which assumes that it is possible to isolate a set of phenomena from all the rest. The existence of symmetry is of great value in constructing explicit examples of effective field theories. In this case the properties of these effective theories are severely constrained by symmetry. In a region of parameter space in which all velocities are much smaller than the speed of light, one can ignore relativity altogether. It is not that there is anything wrong with treating physics in a fully relativistic fashion. It is simply easier not to include relativity if there is no need for that.

An important example to which this discussion applies is a nonrelativistic theory living on a curved $d$-dimensional manifold in the presence of some 
background fields. After integrating out the dynamical degrees of freedom, one can consider the resulting effective action as a functional of the 
background fields. The simplest case to study is a set of two fields - a metric $g_{ij}$ and a $U(1)$ gauge field $A_\mu=(A_0,A_i)$. 
This is the minimal set. In this case, a remarkably simple suggestion made in \cite{SW} is that under diffeomorphisms generated by 
\begin{equation}\label{xi}
\delta t=0\,,\qquad\delta x^i=\xi^i(t,\mathbf{x})
\end{equation}
the background fields transform as
\begin{equation}\label{transNR}
\delta A_0=-\partial_k A_0\xi^k- A_k\dot\xi^k\,,\qquad\delta A_i=-\partial_k A_i\xi^k- A_k\partial_i\xi^k-mg_{ik}\dot\xi^k\,,\qquad
\delta g_{ij}=-\partial_k g_{ij}\xi^k-g_{kj}\partial_i\xi^k-g_{ik}\partial_j\xi^k\,,
\end{equation}
where $\dot{\xi}=\partial_t\xi$ and $m$ is a mass parameter. In addition, the theory is supposed to be invariant under $U(1)$ gauge transformations generated by 
\begin{equation}\label{transG}
\delta A_0=-\dot{\alpha}\,,\qquad
\delta A_i=-\partial_i\alpha\,,\qquad
\delta g_{ij}=0\,.
\end{equation}
Note that the above formulation is universal since there is no need for the explicit transformation rules of the dynamical fields. 

A further development of these ideas has led to the formalism of Newton-Cartan geometry \cite{SonCo} and interesting applications to the Hall liquid. In the later case, symmetry allows one to relate the Hall viscosity with the leading correction to the Hall conductivity (in an expansion in small wave numbers) \cite{HS,Read}. In addition, there are some interesting relations to a theory of gravity with anisotropic scaling in the presence of a $U(1)$ gauge field \cite{Horava,AHH}.

In this paper, we continue our study of the nonrelativistic diffeomorphism invariance \cite{AHH}  with the minimal set of the background fields. In section II, we address three issues related with a 
nonrelativistic limit of some actions in three-dimensional spacetime. First, we consider an additional $U(1)$ gauge field and use the Seiberg-Witten map 
to map one $U(1)$ gauge theory into another. This allows us to effectively construct invariant effective actions within the $\varepsilon$-expansion of \cite{HS}. We then continue by discussing a nonrelativistic limit of the relativistic Klein-Gordon field coupled to a $U(1)$ gauge field. Finally, we 
demonstrate what effective actions can be derived from Chern-Simons gravity. We then go on in Section III to discuss some implications for the viscosity tensor. In particular, we discuss the correction to the Hall viscosity due to a non-uniform magnetic field. In section IV, we present, using the ideas of section II, further examples of invariant actions. On Riemann surfaces of constant non-zero curvature $R$ an effective action is expected to have $1/R$ terms. With this in mind, we construct several examples of such actions. In section V, we consider some issues about spinning particles, focusing 
on the continuity equation. We conclude in section VI with a brief discussion of some further issues related to higher order terms in the $\varepsilon$-expansion, the viscosity tensor, and a redefinition of the background fields. Some technical details are given in the appendices: Appendix A contains the notation and presents a few useful formulas, Appendix B gives an example of the Seiberg-Witten map, Appendix C provides additional information on the viscosity tensor. 

%___________________________________________________________________
\section{Going From $3$ to $2+1$}
\renewcommand{\theequation}{2.\arabic{equation}}
\setcounter{equation}{0}

In this section we will address three issues about a nonrelativistic limit of some actions in three-dimensional spacetime. In our discussion, we 
use the formalism proposed in \cite{SW} for pure gravity. The point there is that the transformation rules \eqref{transNR}-\eqref{transG} 
can be derived by taking a nonrelativistic limit of those for the spacetime metric 
\begin{equation*}
\delta g_{\mu\nu}=-\partial_\lambda g_{\mu\nu}\xi^\lambda-g_{\lambda\nu}\partial_\mu\xi^\lambda-g_{\mu\lambda}\partial_\nu\xi^\lambda
\,,
\end{equation*}
with $x^0=ct$ and ${\mu,\nu}=0,\dots,d$. In this case, for $\xi^\lambda=\bigl(-\frac{\alpha}{mc},\xi^k\bigr)$, the $1/c$ expansion of the metric
is given by 
\begin{equation}\label{metricd+1}
g_{\mu\nu}=
\begin{pmatrix}
-1+{\displaystyle\frac{2A_0}{mc^2}}\,\,&{}\,\,&{\displaystyle\frac{A_i}{mc}}\,\,\\
{}\,\,&{}\,\,&\,\,\\
{\displaystyle\frac{A_i}{mc}}\,\,&{}\,\,& g_{ij}\,\,
\end{pmatrix}\,,
\end{equation}
with $A_0$ and $A_i$ kept fixed as $c\rightarrow\infty$. It is noteworthy that the time component of $\xi^\mu$ becomes a parameter of the gauge transformations \eqref{transG}.

\subsection{Adding $U(1)$ Gauge Field}

Our first goal will be to reexamine this issue in the presence of a $U(1)$ gauge field in three-dimensional spacetime. It transforms under spacetime 
diffeomorphisms as 

\begin{equation}\label{AA}
\delta {\cal A}_\mu=-\partial_\nu{\cal A}_\mu\xi^\nu-{\cal A}_\nu\partial_\mu\xi^\nu\,.
\end{equation}

We consider the $1/c$ expansion of the gauge field

\begin{equation}\label{gaugefield}
	{\cal A}_\mu=\left(\frac{{\cal A}_0}{c},\,\,{\cal A}_i\right)
	\,,
\end{equation}
with ${\cal A}_0$ and ${\cal A}_i$ held fixed as $c\rightarrow\infty$. Then, in the limit $c\rightarrow\infty$, \eqref{AA} becomes

\begin{equation}\label{trans2}
\delta {\cal A}_0=-\partial_k {\cal A}_0\xi^k- {\cal A}_k\dot\xi^k\,,
\qquad
\delta {\cal A}_i=-\partial_k {\cal A}_i\xi^k- {\cal A}_k\partial_i\xi^k
\,.
\end{equation}
In contrast to the case of the spacetime metric, the transformations related with the time component of $\xi^\mu$ decouple in the limit $c\rightarrow\infty$. However, the gauge field ${\cal A}$ still transforms under the gauge transformations in the ordinary way

\begin{equation}\label{transG2}
\delta{\cal A}_0=-\dot{\Lambda}\,,\qquad
\delta{\cal A}_i=-\partial_i\Lambda\,,
%\delta g_{ij}=0\,.
\end{equation}
with $\Lambda$ a parameter.

Since we are dealing with the minimal set of the background fields, we need to determine the relation between the two $U(1)$ gauge theories described by $A$ and $\cal A$. What's needed is a transformation that maps one gauge theory into another. Following Seiberg and Witten \cite{SeW}, we look for a mapping such that 

\begin{equation}\label{SW}
	{\cal A}(A)+\delta_{\epsilon'}{\cal A}(A)={\cal A}(A+\delta_{\epsilon}A)
	\,.
\end{equation}
Here $\delta_{\epsilon'}{\cal A}$ is a shorthand for the set of the transformation laws that contains \eqref{trans2} and \eqref{transG2}. Correspondingly, $\delta_{\epsilon}A$ is that for \eqref{transNR} and \eqref{transG}. $\epsilon'$ and $\epsilon$ stand for the infinitesimal parameters.

In making \eqref{SW} more explicit, we will specialize to the case $d=2$. We start with the nonrelativistic diffeomorphism 
transformations \eqref{transNR} and \eqref{trans2}. 
Like in \cite{SeW}, we write ${\cal A}=A+f(A)$ and then set $\epsilon'=\epsilon=\xi$. Within the $\varepsilon$-expansion\footnote{
Following \cite{HS}, we consider the scaling $A_i\sim\varepsilon^{-1}$, $\partial_i\sim\varepsilon$ and $\partial_t\sim\varepsilon^2$
with everything else, including $A_0$, $m$ and $g_{ij}$, held fixed as $\varepsilon\rightarrow 0$.}, equation \eqref{SW} is solved by 

\begin{equation}\label{A}
	{\cal A}_0=A_0-\frac{1}{2}mv_iv^i+\frac{1}{B}m^2J_i^n v^i\mathsf{d}v_n
	+O(\varepsilon^6)
	\,,\qquad
{\cal A}_i=A_i+mv_i-\frac{1}{B}m^2J^n_i\mathsf{d}v_n+O(\varepsilon^5)
\,,
\end{equation}
where the drift velocity and the complex structure are given by
\begin{equation}\label{drift2}
v^i=-\e^{ij}\frac{E_j}{B}
\,,\qquad
J^n_m=\e_{mk}g^{kn}
\,.
\end{equation}
Here $\mathsf{d}v_i=\dot v_i+v^m\nabla_m v_i$, $B=\e^{ij}\partial_iA_j$, $E_i=\dot A_i-\partial_iA_0$, and $\e^{ij}=\varepsilon^{ij}/\sqrt{g}$ such that $\varepsilon^{12}=1$ and $g=\det g_{ij}$. 

Equation \eqref{SW} generates the change of variables such that ${\cal A}(A)$ obeys the rules \eqref{trans2} which differ from those \eqref{transNR} only by the term proportional to $m$. Hence, it is natural to think of $m$ as a deformation parameter and expand ${\cal A}$ in powers of $m$. In doing so, one 
must respect the fact that the transformation rules are invariant under a scaling operation: $A\rightarrow \lambda A$, $m\rightarrow\lambda m$, 
$g\rightarrow\lambda^0 g$, and ${\cal A}\rightarrow\lambda{\cal A}$. This is the reason for the appearance of negative powers of $B$ in \eqref{A}. It is clear that one can find the mapping at any finite order in $m$. However, it is not clear whether the perturbative series is convergent or not. Note that the expansion in powers of $m$ is consistent with the $\varepsilon$-expansion. This follows from the form of \eqref{transNR}. Indeed, the only term, which changes the order of terms in the $\varepsilon$-expansion of an effective action, is proportional to $m$. 

Interestingly, the equation can be solved explicitly for the inverse mapping. In this case, its solution is simply

\begin{equation}\label{A-}
	A_0={\cal A}_0+\frac{1}{2}m{\cal V}_i{\cal V}^i
	\,,\qquad
	A_i={\cal A}_i-m{\cal V}_i
	\,,\qquad
	{\cal V}_i=-\e_{ij}\frac{{\cal E}^j}{{\cal B}}
\,,
\end{equation}
where ${\cal B}=\e^{ij}\partial_i{\cal A}_j$, ${\cal E}_i=\dot{\cal A}_i-\partial_i{\cal A}_0$.

Now we are ready to extend the analysis to the gauge transformations \eqref{transG} and \eqref{transG2}. From \eqref{SW} and \eqref{A}, it immediately 
follows that 

\begin{equation}\label{2U(1)}
	\Lambda=\alpha\,.
\end{equation}

We conclude our discussion of the mapping by giving a few examples of effective actions verifying and illustrating the construction. As before, we 
focus on the case $d=2$.

The first example is the Chern-Simons action. It is invariant under the spacetime diffeomorphisms and $U(1)$ gauge transformations. Using 
\eqref{gaugefield} and \eqref{A}, we get\footnote{We label a set of invariant actions for the use in section VI.}

\begin{equation}\label{CS}
	\begin{split}
	S_1=\lim_{c\rightarrow\infty}\int d^3x \,\,\varepsilon^{\mu\nu\lambda}{\cal A}_\mu\partial_\nu{\cal A}_\lambda
	=&
	\int dt d^2x\sqrt{g}\,\biggl[{\cal A}_0{\cal B}-\e^{ij}{\cal A}_i{\cal E}_j\biggr]\\
	=&
	\int dt d^2x\sqrt{g}\,\biggl[
	\frac{\varepsilon^{\mu\nu\lambda}}{\sqrt{g}}
	A_\mu\partial_\nu A_\lambda
	+m\Bigl(B+\frac{1}{2}m\Omega\Bigr)v_iv^i
	-m^2\e^{ij}v_i\mathsf{d}v_j \biggr]\,\,\,\,+o(\varepsilon^4)
\,,
\end{split}
\end{equation}
where we have introduced the vorticity $\Omega=\e^{ij}\partial_i v_j$. The object $\varepsilon^{\mu\nu\lambda}$ is a completely antisymmetric symbol 
with $\varepsilon^{012}=1$. The action on the right hand side is invariant under both the nonrelativistic diffeomorphisms (up to order $\varepsilon^4$) and the gauge transformations.

The next example is the Yang-Mills action. On dimensional grounds, we take the coupling $g^2$ to be proportional to $mc$. With \eqref{metricd+1}, \eqref{gaugefield} and \eqref{A}, we find

\begin{equation}\label{BB}
   S_2=\lim_{c\rightarrow\infty} \frac{1}{2}\int d^3x\sqrt{g^{(3)}}\,\frac{1}{mc}{\cal F}_{\mu\nu}{\cal F}^{\mu\nu}
=\int dtd^2x\sqrt{g}\,\frac{1}{m}{\cal B}^2
	=\int dtd^2x\sqrt{g}\,\,\biggl[\frac{1}{m}(B+m\Omega)^2
	+2mB\nabla^n\Bigl(\frac{\mathsf{d}v_n}{B}\Bigr)
	\biggr]\,\,+o(\varepsilon^4)
	\,,
\end{equation}

\noindent where $g^{(3)}=\det g_{\mu\nu}$ and ${\cal F}_{\mu\nu}=\partial_\mu{\cal A}_\nu-\partial_\nu{\cal A}_\mu$. Our general discussion above shows that the resulting action is invariant up to the 4th order in $\varepsilon$. Of course, one can verify this explicitly by using \eqref{transNR}.

To find the relativistic counterpart of the Wen-Zee action \cite{WZ}, we following \cite{GRS} introduce the following gauge invariant fields

\begin{equation}\label{bu}
b=\sqrt{\frac{1}{2}{\cal F}_{\mu\nu}{\cal F}^{\mu\nu}}\,,\qquad
u^\mu=\frac{1}{2b}\e^{\mu\nu\lambda}{\cal F}_{\nu\lambda}\,,
\end{equation}
where $\e^{\mu\nu\lambda}$ is a completely antisymmetric tensor. A simple but somewhat lengthy algebra shows that\footnote{The first term on the left hand side of \eqref{WZ} is gauge invariant as it follows from \cite{GRS}.}

\begin{equation}\label{WZ}
	\begin{split}
S_3=\lim_{c\rightarrow\infty}\frac{1}{2}&\int d^3x\sqrt{g^{(3)}}\,
\left[\e^{\mu\nu\lambda}\e^{\alpha\beta\gamma}{\cal A}_\mu u_\alpha
\Bigl(\nabla_\nu u_\beta \nabla_\lambda u_\gamma -
\frac{1}{2}R^{(3)}_{\nu\lambda\beta\gamma}\Bigr)
- b\,\e^{\mu\nu\lambda}u_\mu\nabla_\nu u_\lambda\right]\\
=&\int dtd^2x\sqrt{g}
\biggl[\frac{\varepsilon^{\mu\nu\lambda}}{\sqrt{g}}\omega_\mu\partial_\nu {\cal A}_\lambda+\frac{1}{2}{\cal B}{\cal O}\biggr]
\\
=&
\int dtd^2x\sqrt{g}
\biggl[\frac{\varepsilon^{\mu\nu\lambda}}{\sqrt{g}}\omega_\mu\partial_\nu A_\lambda
+\frac{1}{2}\Bigl(B+m\Omega\Bigr)\Omega
+m\Bigl(\omega_0+\omega_iv^i\Bigr)\Omega
+\frac{1}{2}mB\nabla^n\Bigl(\frac{\mathsf{d}v_n}{B}\Bigr)
-m\e^{ij}\omega_i\mathsf{d}v_j\biggr]\,\,\,\,+o(\varepsilon^4)
\,,
\end{split}
\end{equation}
with $\omega_\mu=(\omega_0,\omega_i)$ such that 
\begin{equation}\label{omega}
	\omega_0=\frac{1}{2}\varepsilon^{ab}e^j_a\dot e^b_j\,,\qquad
	\omega_i=\frac{1}{2}\varepsilon^{ab}e^j_a\nabla_i e_j^b
	\,.
\end{equation}
In this example, $R^{(3)}_{\nu\lambda\beta\gamma}$ is a three-dimensional Riemann tensor, $\omega_i$ is a minimal spin connection in two-dimensions, 
$e_i^a$ is a zweibein, and ${\cal O}$ is a vorticity of ${\cal V}$, ${\cal O}=\e^{ij}\partial_i{\cal V}_j$. In the first step, we 
used\footnote{Interestingly, this demonstrates explicitly that the different relativistic actions (for instance, \eqref{BB} and \eqref{b}) may have the same nonrelativistic limit.}  

\begin{equation}\label{b}
\lim_{c\rightarrow\infty}\int d^3x\sqrt{g^{(3)}} \,b\,\e^{\mu\nu\lambda}u_\mu\nabla_\nu u_\lambda=
-\frac{1}{m}\int dtd^2x\sqrt{g}\,{\cal B}^2\,.
\end{equation}
Apart from the obvious fact that the right hand side of \eqref{WZ} is invariant under the $U(1)$ gauge transformations, it is also invariant under the local $SO(2)$ rotations of $\omega_0$ and $\omega_i$. One can easily check the last statement by using the definition of $\Omega$ together with an integration by parts.

Finally, to complete the picture, we give the relativistic counterparts of the two remaining contributions to the nonrelativistic action of \cite{HS} which are invariant up to order $\varepsilon^4$

\begin{equation}\label{R}
S_4=\lim_{c\rightarrow\infty}\int d^3x\sqrt{g^{(3)}}\,\frac{1}{mc}R^{(3)}b=
\int dtd^2x\sqrt{g}\,\frac{1}{m}R{\cal B}=
\int dtd^2x\sqrt{g}\,\frac{1}{m}R
\Bigl[B+m\Omega\Bigr]\,\,\,\,+o(\varepsilon^4)
\,,
\end{equation}

\begin{equation}\label{dBdB}
	\begin{split}
S_5=\lim_{c\rightarrow\infty}\int d^3x\sqrt{g^{(3)}}\,\frac{1}{mc}g^{\mu\nu}b^{-1}\partial_\mu b\,\partial_\nu b
=&\int dtd^2x\sqrt{g}\,\frac{1}{m}g^{ij}{\cal B}^{-1}\partial_i{\cal B}\,\partial_j{\cal B} \\
	=&\int dtd^2x\sqrt{g}\,\frac{1}{m}g^{ij}B^{-1}
	\biggl[\biggl(1-\frac{m}{B}\Omega\biggr)
	\partial_iB\partial_jB 
	+2m\partial_i\Omega\partial_jB\biggr]
	\,\,\,\,+o(\varepsilon^4)
	\,.
\end{split}
\end{equation}	
	
We close this section with the following remarks:

(i) In \cite{AHH}, we considered a symmetry group that includes nonrelativistic diffeomorphisms and a product of two gauge groups $U(1)\times SO(d)$. In that case, however, we didn't map one gauge group into another with the help of the Seiberg-Witten map.

(ii) It is surprising that the solution \eqref{A} of the Seiberg-Witten equations contains $v^i$ which is a good approximation for 
describing the drift velocity of the Hall fluid as well as the material derivative ${\mathsf d}$ which is known from the Euler equations of fluid 
dynamics.

(iii) It is interesting to ask what would happen if another form of mapping was used. To answer this question, let us consider a simple modification of \eqref{A-} that provides such a form. In doing so, a useful fact is that if ${\cal Q}_0$ and ${\cal Q}_i$ transform as ${\cal A}_0$ and ${\cal A}_i$ in \eqref{trans2}, then ${\cal A}_0+{\cal Q}_0$ and ${\cal A}_i+{\cal Q}_i$ obey the transformation rules \eqref{transNR}. Thus, \eqref{A-} can be modified as 

\begin{equation}\label{A-m}
A_0={\cal A}_0+\frac{1}{2}m{\cal V}_i{\cal V}^i+{\cal Q}_0
	\,,\qquad
	A_i={\cal A}_i-m{\cal V}_i+{\cal Q}_i
	\,,
\end{equation}
with gauge invariant ${\cal Q}_0$ and ${\cal Q}_i$. In a generic case, the ${\cal Q}$'s are constructed from both the sets of the gauge fields $A_\mu$ and ${\cal A}_\mu$. For given ${\cal Q}_0$ and ${\cal Q}_i$, equation \eqref{A-m} can be solved, at least perturbatively. A solution provides a new mapping, let us call it ${\cal A}'(A)$. By using it, we can construct a new set of invariant actions. These actions will, in general, differ from those we presented above. The point, however, is that the effective actions are related by a field redefinition. As an important illustration of these ideas, we will consider a simple example in Appendix B.

(iv) The examples of effective actions invariant up to order $\varepsilon^2$ were discussed in \cite{HS,AHH}. The point here is that the proposed method allows one to efficiently construct invariant actions, at {\it any} finite order in $\varepsilon$, by taking a nonrelativistic limit of their relativistic counterparts and solving equation \eqref{SW} at that order. In \cite{AHH}, we gave the examples of how it works for pure gravity, where the mapping $U(1)\rightarrow U(1)$ is not needed. In the present paper, we have extended the construction to include an additional $U(1)$ gauge field. Note that the derivative corrections have been recently studied in \cite{AG}. In contrast, we don't restrict ourselves to the terms quadratic in the background fields as well as to the fourth order in the gradient expansion.

%____________________________________________________________________________________________________
\subsection{Klein-Gordon Field Coupled to $U(1)$ Gauge Field}

Now we will address the issue of a nonrelativistic limit of the relativistic theory of a complex scalar field. The main point is to understand what is going on in the presence of a $U(1)$ gauge field. To this end, we use the expansion \eqref{gaugefield} which is regular in the $c\rightarrow\infty$ limit.\footnote{See \cite{JK}, for an example of singular expansions.} 

The relativistic action is given by 

\begin{equation}\label{KG}
S=\frac12\int d^3x\sqrt{g^{(3)}}\,\Bigl[g^{\mu\nu}D_\mu\Phi^\dagger D_\nu\Phi+(mc)^2\Phi^\dagger\Phi
\Bigr]\,,
\end{equation}
where $D_\mu\Phi=\partial_\mu\Phi+i{\cal A}_\mu\Phi$ is the gauge covariant derivative. The action is invariant under the $U(1)$ gauge transformations
$\delta{\cal A}_\mu=-\partial_\mu\Lambda$ and $\delta\Phi=i\Lambda\Phi$.

First, we take the $c\rightarrow\infty$ limit by using the expansions \eqref{metricd+1} and \eqref{gaugefield} complemented by an 
expansion for the scalar field $\Phi=\e^{imc^2t}\psi/\sqrt{mc}$. Doing so, we find 

\begin{equation}\label{Schr}
S=\frac12 \int dtd\mathbf{x}\sqrt{g}
\biggl[i\psi^\dagger\overset\leftrightarrow{\partial}_t\psi+2({\cal A}_0-A_0)\psi^\dagger\psi+
\frac{g^{ij}}{m}
\Bigl(\partial_i\psi^\dagger-i(A_i-{\cal A}_i)\psi^\dagger\Bigr)
\Bigl(\partial_j\psi+i(A_j-{\cal A}_j)\psi\Bigr)
\biggr]	
\,,
\end{equation}
where $\psi^\dagger\overset\leftrightarrow{\partial}_t\psi=\psi^\dagger\dot\psi-\dot\psi^\dagger\psi$. 

Using the mapping \eqref{A}, we can write this in the form 

\begin{equation}\label{Schr-inv}
S=\frac12 \int dtd\mathbf{x}\sqrt{g}\biggl[i\psi^\dagger\overset\leftrightarrow{\partial}_t\psi-mV_iV^i\psi^\dagger\psi+
\frac{g^{ij}}{m}
\Bigl(\partial_i\psi^\dagger+imV_i\psi^\dagger\Bigr)
\Bigl(\partial_j\psi-imV_j\psi\Bigr)
\biggr]	
\,,
\end{equation}
with
\begin{equation}\label{drift}
	V_i={\cal V}_i=v_i-\frac{1}{B}mJ^n_i{\mathsf{d}}v_n+\dots
	\,\,.
	\end{equation}
This form is notable for its relation to the description of a nonrelativistic system of noninteracting particles in the background geometry 

\begin{equation}\label{foliation-geometry}
	ds^2=-c^2dt^2+g_{ij}\bigl(dx^i-V^idt\bigr)\bigl(dx^j-V^j dt\bigr)
	\,,
\end{equation}
with $V^i$ a shift vector. The spacetime metric \eqref{foliation-geometry} naturally appears in studying a subgroup of foliation-preserving diffeomorphisms \eqref{xi}, with the time coordinate $x^0$ chosen to be a global time. In this case the spatial metric and gauge field transform as 
in \eqref{transNR}. 

One way to turn the background gauge field in \eqref{Schr-inv} on is to replace the derivatives as follows: $\partial_t\rightarrow\partial_t-iA_0^{\text{elm}}$ and $\partial_i\rightarrow\partial_i-iA_i^{\text{elm}}$, where $A^{\text{elm}}$ is a usual $U(1)$ gauge field transforming as a one-form, and then define the new fields $A_0=A_0^{\text{elm}}+\frac{1}{2}mV_iV^i$ and $A_i=A_i^{\text{elm}}-mV_i$ \cite{AHH}. 
This results in the minimally coupled gauge field $A_\mu$. Note that in this case $\psi$ transforms under the gauge transformations as $\delta\psi=-i\alpha\psi$, with a parameter $\alpha$ coming from a time component of the parameter $\xi^\mu$ of spacetime diffeomorphisms.

In our present discussion, however, we have found another way that is to express the shift vector in terms of the background gauge field. The resulting 
description is non-minimal coupling, with $\psi$ a gauge singlet. The latter follows from the transformation rules under the two $U(1)$'s combined with 
\eqref{2U(1)}, namely $\delta\psi=-i(\alpha-\Lambda)\psi=0$. Of course, it is also clear from \eqref{Schr-inv} and \eqref{drift}, directly.

\subsection{Chern-Simons Gravity}

The last issue concerning a nonrelativistic limit of three-dimensional actions that we will briefly discuss here is what effective actions can be derived from Chern-Simons gravity \cite{EW}.

First, we consider the minimal spin connection. In this case, the fundamental variable is a dreibein $\e_\mu^A$. The spin connection is given 
by ${\text w}_\mu^{AB}=\eta^{BC}\e^A_\nu\nabla_\mu\e^\nu_C$, with the Minkowski metric $\eta_{AB}=\text{diag}(-1,1,1)$. The gravitational Chern-Simons action 
can be written in the following form

\begin{equation}\label{CS-gravity}
	S=\int d^3x\,\varepsilon^{\mu\nu\lambda}\,{\text w}^A_\mu
	\biggl(\partial_\nu{\text w}^A_\lambda+
	\frac{1}{3}\varepsilon_{ABC}\,{\text w}_\nu^B{\text w}^C_\lambda
	\biggr) 
	\,,
\end{equation}
where ${\text w}^A_\mu=\frac{1}{2}\eta^{AD}\varepsilon_{DBC}\,{\text w}_\mu^{BC}$.

Using the $1/c$ expansion of the minimal spin connection 

\begin{equation}\label{spin-c}
{\text w}^0_0=\frac{1}{c}\omega^0_0=\frac{1}{c}\Bigl(\omega_0-\frac{1}{2m}B\Bigr)\,,
\qquad
{\text w}^0_i=\omega_i^0=\omega_i\,,
\qquad
{\text w}^a_\mu=0
\,
\end{equation}
and taking $c\rightarrow\infty$, we obtain

\begin{equation}\label{wdw}
	S=\int dtd^2x\,\varepsilon^{\mu\nu\lambda}\omega_\mu^0\partial_\nu\omega_\lambda^0
	=\int dtd^2x
	\biggl[\varepsilon^{\mu\nu\lambda}\omega_\mu\partial_\nu\omega_\lambda-\frac{1}{2m}\sqrt{g}RB
	\biggr]
\,.
\end{equation}
This action is invariant under both the nonrelativistic diffeomorphisms \eqref{transNR} and the gauge transformations \eqref{transG}. In \cite{AHH}, we 
derived it from another invariant action by changing variables. 

In addition to the action \eqref{CS-gravity}, there is another one \cite{EW}

\begin{equation}\label{CS-gravity2}
	S=\int d^3x\,\varepsilon^{\mu\nu\lambda}\,
	\Lambda\,\e^A_\mu
	\biggl(\partial_\nu\e^A_\lambda+
	\varepsilon_{ABC}\,{\text w}_\nu^B\e^C_\lambda
	\biggr) 
	\,,
\end{equation}
where $\Lambda$ is a scale parameter of dimension $[\text{momentum}]^2$. In terms of Lie-algebra-valued forms, it is given by 
$\int\text{Tr}(e \wedge T)$, with $T$ the torsion 2-form. Thus, if we choose ${\text w}^A_\mu$ to be the minimal spin connection, then 
this action vanishes identically.

Now let us look at \eqref{CS-gravity2} in the $c\rightarrow$ limit. It seems appropriate to simply set $\Lambda=(mc)^2$ but this leads to 
trouble. One can easily check by using the formulas \eqref{e} that the coefficient in front of the first term is infinite.\footnote{Interestingly, such a term, also with a divergent coefficient, appears in an effective action after integrating the massive fermions out \cite{fradkin}.} What happens if, like in section A, there is an additional gauge field ${\cal A}_\mu$. In this case, the situation is better than before. If we let 
$\Lambda=b$, with $b$ given by \eqref{bu}, then we find that in the $c\rightarrow\infty$ limit the first term is finite

\begin{equation}\label{Hall-v}
	S=2\int dtd^2x\,\sqrt{g}\,{\cal B}
	\biggl[\omega_0-\frac{1}{m}\omega_iA^i-\frac{1}{2m}B\biggr]
	\,.
\end{equation}
It is invariant under nonrelativistic diffeomorphisms because $\omega_0-\frac{1}{m}\omega_iA^i-\frac{1}{2m}B$ is a scalar \cite{AHH}. However, the invariance under the $U(1)$ as well as $SO(2)$ gauge transformations is lost. It is restored by the second term. In this work, we won't go much into details aside from saying a few words in the next section.

%____________________________________________________________________________
\section{Implications for Viscosity Tensor}
\renewcommand{\theequation}{3.\arabic{equation}}
\setcounter{equation}{0}

It is known \cite{read} that the Wen-Zee action gives rise to the Hall viscosity $\eta^{\text{\tiny H}}$ \cite{zograf}. In flat space it is defined as a coefficient in front of an antisymmetric part of the viscosity tensor. Now let us discuss how the above is modified if the invariance under nonrelativistic diffeomorphisms is imposed. We consider the case that actions are invariant up to order $\varepsilon^4$.

We begin with the action \eqref{WZ}. Because the leading term of its $\varepsilon$-expansion coincides with the Wen-Zee action, one can think off it as an invariant (under nonrelativistic diffeomorphisms) extension of the Wen-Zee action. It is convenient to write the action as\footnote{In this section, we use the notation of \cite{HS} for the prefactors.}

\begin{equation}\label{WZ1}
	S=\frac{\kappa}{2\pi}
	\int dtd^2x\sqrt{g}
	\biggl[{\cal B}\,\omega_0+\e^{ij}{\cal E}_i\omega_j+\frac{1}{2m}{\cal B}\bigr({\cal B}-B\bigl)\biggr]
\,.
\end{equation}

We will be essentially interested in the linear response to a weak time-dependent perturbation of the metric $g_{ij}(t)$. Following \cite{fradkin}, we 
write $e^a_i=\delta^a_i+u^a_i(t)$. In the problem at hand, we can forget about the difference between tangent $(a)$ and local coordinate $(i)$ indices and freely switch between those. The indices are raised and lowered by means of the Euclidean metric $\delta_{ij}$. For the sake of simplicity, 
we assume that the tensor $u_{ij}$ is symmetric. Then $u_{ij}$ and $\dot u_{ij}$ are the strain and strain-rate tensors, respectively. In this framework
the metric is given by $g_{ij}=\delta_{ij}+ 2u_{ij}+u_{ik}u_{kj}$. Moreover, the minimal spin connection $\omega_i$ vanishes and, as a consequence, 
\eqref{WZ1} reduces to  

\begin{equation}\label{WZ2}
	S=\frac{\kappa}{2\pi}
	\int dtd^2x\,\varepsilon^{ij}\partial_i{\cal A}_j
	\Bigl[\omega_0+\frac{1}{2m}\bigr({\cal B}-B\bigl)\Bigr]
\,.
\end{equation}

The part of the action involving $u\dot u$ can be easily found. We have

\begin{equation}\label{m-current}
	S=\frac{\kappa}{4\pi}\int dtd^2x\,
	\biggl[
	\Bigl(
	\mathsf{B}+m\mathsf{\Omega}-
	2\frac{m}{\mathsf{B}}\varepsilon_{nl}\partial_n\mathsf{B}\mathsf{v}_l
\Bigr)
	\varepsilon_{ij}u_{jk}\dot u_{ki}+
	\mathsf{B}\,\mathsf{t}^{1}_{ij}
	\Bigl(u_{ik}\dot u_{kj}-{\text {tr}}u\,\dot u _{ij}
	\Bigr)
	\biggr]
	\,.
\end{equation}
Here $t^1_{ij}$ is a symmetric tensor defined in Appendix C. We use letters in a sans serif font to denote the fields in flat space (at $g_{ij}=\delta_{ij}$). In the problem under consideration, the stress tensor response, linear in $\dot u$, can be read from the result of variation $-\delta S/\delta u_{ij}$. After a short calculation, we obtain

\begin{equation}\label{T}
	T_{ij}=\frac{\kappa}{4\pi}
	\biggl[
	\Bigl(
	\mathsf{B}+m\mathsf{\Omega}-
	2\frac{m}{\mathsf{B}}\varepsilon_{nl}\partial_n\mathsf{B}\mathsf{v}_l
	\Bigr)
	\bigl(\varepsilon_{ik}\dot u_{jk}+\varepsilon_{jk}\dot u_{ik}\bigr)+
	\mathsf{B}\bigl(\mathsf{t}^{1}_{nm}\dot u_{nm}\delta_{ij}-\mathsf{t}^{1}_{ij}{\text {tr}}\dot u\bigr)
	\biggr]
	\,.
\end{equation}
As a result, we learn that the non-zero coefficients in the antisymmetric part of the viscosity tensor are given by 

\begin{equation}\label{Hall}
\eta^{\text{\tiny H}}(\kappa)=\frac{\kappa}{4\pi}
\Bigl(
\mathsf{B}+m\mathsf{\Omega}-
2\frac{m}{\mathsf{B}}\varepsilon_{nl}\partial_n\mathsf{B}\mathsf{v}_l
\Bigr)
\,,\qquad
\eta^{\text{\tiny V}}_1(\kappa)=\frac{\kappa}{2\pi}\mathsf{B}
\,,
\end{equation}
where the $\eta$'s are defined in Appendix C. The point here is that in the Hall viscosity, the two terms $\mathsf{\Omega}$ and 
$\partial\mathsf{B}\,\mathsf{v}$ combine to give a $\varepsilon^2$-correction to the well-known leading order term. 
The later is a correction due to a non-uniform magnetic field.

We will now carry out a precisely analogous computation for the actions \eqref{CS} and \eqref{BB} which are the corresponding generalizations of the Chern-Simons and Yang-Mills actions. To order $\varepsilon^2$ there are no contributions to the viscosity tensor but this is no longer true at the next order. The relevant terms are  

\begin{align}\label{CSBB}
	S=&\frac{\nu}{4\pi}\int dtd^2x\sqrt{g}
	\Bigl[{\cal A}_0{\cal B}-\e^{ij}{\cal A}_i{\cal E}_j\Bigr]=\frac{\nu m^2}{4\pi}\int dtd^2x\,\varepsilon^{ij}v_j{\mathsf d}v_i+\dots
	\,\,,\\
	S=&-\frac{\epsilon}{4\pi m}\int dtd^2x\sqrt{g}\,{\cal B}^2=-\frac{\epsilon m}{2\pi}\int dtd^2x\sqrt{g}\,B\nabla^i\Bigl(\frac{{\mathsf d}v_i}{B}
	\Bigr)+\dots\,\,.
\end{align}
From those, we obtain the non-zero coefficients of the viscosity tensor

\begin{equation}\label{Hall2}
\eta^{\text{\tiny H}}_2(\nu)=\frac{\nu}{2\pi}\mathsf{B}
\,,\qquad
\eta^{\text{\tiny H}}(\epsilon)=\frac{\epsilon}{\pi}
\frac{m}{\mathsf{B}}\varepsilon_{nl}\partial_n\mathsf{B}\mathsf{v}_l
\,,\qquad
\eta^{\text{\tiny V}}_1(\epsilon)=-\frac{\epsilon}{\pi}\mathsf{B}
\,.
\end{equation}
As before, there is the $\varepsilon^2$-correction to the Hall viscosity due to a non-uniform magnetic field.

We can similarly analyze the remaining actions \eqref{R}, \eqref{dBdB}, and \eqref{wdw}. In this case, a short check shows that there are no relevant 
terms that would give rise to the viscosity tensor at order $\varepsilon^2$.

At this point a few short comments are in order:

(i) The realization of the invariant effective action via the $\varepsilon$-expansion leads to 
that for the stress tensor. From this point of view it is not surprising that the Hall viscosity can be written as a series in "$\varepsilon$". It is infinite. The reason for this last statement is that in \eqref{WZ2} the first term gives rise to a contribution $\tfrac{\kappa}{4\pi}{\cal B}$ to the viscosity. When rewritten in terms of the original fields, one has $\tfrac{\kappa}{4\pi}\left(\mathsf{B}+m\mathsf{\Omega}+m^2\partial_n(\mathsf{d}\mathsf{v}_n/\mathsf{B})+\dots \right)$.\footnote{It becomes a lot of work to find higher order terms in \eqref{A} and we restrict ourselves to a few leading terms.} 

(ii) The above formulas give the expressions for the Hall viscosity up to order $\varepsilon^2$. However, this is not the only point to learn from our 
analysis. At higher orders, the structure of the viscosity tensor gets more involved because of a non-uniform magnetic field and a non-zero drift velocity. In particular, as shown in Appendix C, already at order $\varepsilon^2$ it includes $23$ free parameters. If a physical system is invariant under the group 
of nonrelativistic diffeomorphisms, then one would expect less number of parameters because of symmetry restrictions. The given examples show how this works for the actions invariant up to order $\varepsilon^4$. 

(iii) For small frequencies and wave numbers $\vec q$ invariance under the group of nonrelativistic diffeomorphisms allows one to derive the relation between the leading contribution to the Hall viscosity $\eta^{\text{\tiny H}}$ (the first term in \eqref{Hall}) and the $q^2$ part of the Hall conductivity \cite{HS}. The form of the $\varepsilon^2$-correction to $\eta^{\text{\tiny H}}$ assumes that the response to external electromagnetic field perturbations around a constant magnetic field has to be computed beyond the linear approximation. 

Our remaining goal will be to discuss some issues that arise in the first order formalism for gravity. In this case, the metric and the spin connection are supposed to be independent. Thus, in addition to the stress tensor defined as the response of the action to a change in the metric, one may consider a new object called the spin current and defined as the response of the action to a change in the spin connection. At this point, one might ask what is the significance of using the first order formalism for the Hall viscosity. For this discussion, we will be more specific and consider what is perhaps one of the best understood examples in Chern-Simons gravity \cite{EW}, namely the action \eqref{CS-gravity2}. In the case of the minimal (torsion-free) spin connection expressed in terms of the dreibein, this action vanishes identically. This is not the case anymore in the first order formalism. 

To determine the linear response to a weak time-dependent perturbation of the two-dimensional metric (zweibein), first we have to take the limit $c\rightarrow\infty$ in \eqref{CS-gravity2} with $\Lambda=b$. It is particularly simple for the first term which is independent of the spin connection. The result is given by \eqref{Hall-v}. In the case of interest, it becomes\footnote{Like in \eqref{WZ1}, we introduce a prefactor.}  

\begin{equation}\label{Hall-v2}
	S=\frac{\kappa'}{2\pi}
	\int dtd^2x\,\varepsilon^{ij}\partial_i{\cal A}_j
	\Bigl[\omega_0-\frac{1}{2m}B\Bigr]
\,.
\end{equation}
The above action is manifestly invariant under the $U(1)$ gauge transformations. To make it invariant under the $SO(2)$ gauge transformations, we 
make a simple addition to the action 

\begin{equation}\label{Hall-v3}
	S\rightarrow S+\frac{\kappa'}{2\pi}
	\int dtd^2x\,\varepsilon^{ij}\partial_j{\cal A}_i\,\omega_0^0
\,,
\end{equation}
where $\omega^0_0$ is a non-minimal spin connection defined by $\text{w}^0_0=\omega^0_0/c$. It follows from \eqref{spin-c} that if $\omega^0_0$ is the 
minimal spin connection, then \eqref{Hall-v3} vanishes, as it must.

Having derived the action, we can calculate the bilinear terms $u\dot u$ and, in the fashion analogous to \eqref{T}, obtain 

\begin{equation}\label{Hall3}
\eta^{\text{\tiny H}}(\kappa')=
\frac{\kappa'}{4\pi}
\Bigl(\mathsf{B}+m\mathsf{\Omega}+
2\frac{m}{\mathsf{B}}\varepsilon_{nl}\partial_n\mathsf{B}\mathsf{v}_l
\Bigr)
\,,\qquad
\eta^{\text{\tiny V}}_1(\kappa')=-\frac{\kappa'}{2\pi}\mathsf{B}
\,.
\end{equation}
We have assumed that $\omega^0_0$ scales as $\varepsilon^2$. In this case, the addition \eqref{Hall-v3} doesn't matter for calculations at order $\varepsilon^2$. It becomes relevant for $\varepsilon^4$. If $\omega^0_0$ scales as $\varepsilon^0$, then one must add its contribution to the viscosity coefficients. Both kinds of scaling seem plausible as it follows from the expression for the minimal spin connection \eqref{spin-c}. It is worth noting that the gravitational Chern-Simons action \eqref{CS-gravity2} gives rise to the Hall viscosity whose explicit expression is quite similar to that derived from the ordinary Chern-Simons action \eqref{WZ1}.

The physical interpretation of $\omega^0_0$ is not clear. There is a good deal of literature on defects and geometries with torsion in condensed matter physics\footnote{See, e.g., \cite{kleinert} and references therein.} but the issue of time-dependent two-dimensional 
geometries with torsion and their possible extensions by adding a "time-component" of the non-minimal spin connection deserves to be addressed more thoroughly.

\section{More Examples of Invariant Actions}
\renewcommand{\theequation}{4.\arabic{equation}}
\setcounter{equation}{0}

So far on the way to our goal of constructing invariant actions, we have followed a line of thought that nonrelativistic diffeomorphisms are a subgroup of foliation-preserving diffeomorphisms, with the time coordinate $x^0$ chosen to be a global time. In practice, this means that the actions can be constructed in two steps. The first step is to take an invariant action in three-dimensional spacetime and then find its nonrelativistic limit. The resulting action is a functional of $g$, $\omega$, and ${\cal A}$. Thus, the second step is to change the variables: ${\cal A}\rightarrow A$. At this point, a natural question to ask is whether we can find ways to cut corners. One way to do so is to begin by constructing an invariant action in terms of $g$, $\omega$, and ${\cal A}$.
 
As an important illustration of this idea, we will now present a few examples of invariant actions. Since the minimal spin connection $\omega^0_\mu=(\omega^0_0,\omega_i)$ transforms under nonrelativistic diffeomorphisms like the gauge field ${\cal A}$, it is more convenient to use it instead of using $(\omega_0,\omega_i)$. 

We begin with what is called the geometric action. The idea is to combine the Chern-Simons action with that of 
Wen-Zee as $(A+s\omega)d(A+s\omega)$, where $s$ is a parameter ("spin").\footnote{See, e.g., \cite{Haldane}.} The point is that such defined action is not invariant under nonrelativistic diffeomorphisms. Given the minimal spin connection and gauge field, our proposal for the invariant action is that 

\begin{equation}\label{GA}
	S=\int dtd^2x\,\varepsilon^{\mu\nu\lambda}\bigl({\cal A}_\mu+s\omega^0_\mu\bigr)\partial_\nu
	\bigl({\cal A}_\lambda+s\omega^0_\lambda\bigr)
	\,.
\end{equation}
As a check, note that it includes the expected term $(A+s\omega)d(A+s\omega)$. Explicitly,

\begin{equation}\label{GA2}
\begin{split}
S=	\int dtd^2x\sqrt{g}
\biggl[&
\frac{\varepsilon^{\mu\nu\lambda}}{\sqrt{g}}\bigl(A_\mu+s\omega_\mu\bigr)\partial_\nu
	\bigl(A_\lambda+s\omega_\lambda\bigr)-\frac{s}{2m}B\Bigl(B+\frac{s}{2}R\Bigr)\\
	&+mv_iv^i\Bigl(B-\frac{s}{2}R+\frac{m}{2}\Omega\Bigr)
	+2s\frac{m}{B}g^{ij}E_i\tilde{\cal E}_j+\frac{m}{B}g^{ij}(mE_i+s\partial_iB){\mathsf d}v_j
	\biggr]\,\,\,\,+o(\varepsilon^4)
	\,.
\end{split}
\end{equation}
The remaining terms are required for invariance under nonrelativistic diffeomorphisms. Here $R=2\e^{ij}\partial_i\omega_j$. We have also defined the electric field with respect to the minimal spin connection $\tilde{\cal E}_i=\dot\omega_i-\partial_i\omega^0_0$.

So far we have considered the effective actions compatible with the simplest background: a uniform magnetic field and Euclidean geometry. What happens if we replace Euclidean geometry by that of a more general compact Riemann surface? According to the uniformization theorem, simply connected Riemann surfaces are classified as elliptic (positively curved), parabolic (flat), and hyperbolic (negatively curved). In particular it admits a metric of constant curvature. The natural intuitive answer is that we should extend our analysis to the elliptic and hyperbolic cases. In this paper we will concentrate on one aspect of the extension: a $1/R$ expansion. This is a crucial distinction from the parabolic case. A typical physical situation to which our formalism might be applicable is that of electrons being placed on a sphere. Here one can study, for instance, a response to curvature \cite{W,AG-w}. 

Since both $R$ and ${\cal B}$, being gauge invariant, are the scalars under nonrelativistic diffeomorphisms, the vectors can be obtained by simply 
applying spatial derivatives. This is a good starting point for constructing effective actions, like \eqref{R} and \eqref{dBdB}. From the latter case, 
two further suggestions come to mind: 

\begin{equation}\label{R1}
	S=\int dtd^2x\sqrt{g}\,\frac{1}{m}g^{ij}R^{-1}\partial_i{\cal B}\,\partial_j{\cal B}=\int dtd^2x\sqrt{g}\,\frac{1}{m}g^{ij}R^{-1}
\biggl[\partial_iB\partial_jB+
2m\partial_iB\partial_j
\Bigl(\Omega+\nabla^n\Bigl(\frac{\mathsf{d}v_n}{B}\Bigr)\Bigr)
	+m^2\partial_i\Omega\partial_j\Omega\biggr]\,\,\,\,+o(\varepsilon^4)
	\,
\end{equation}
and 	
\begin{equation}\label{R2}
	S=\int dtd^2x\sqrt{g}\,\frac{1}{m}g^{ij}R^{-1}\partial_i R\,\partial_j{\cal B}=\int dtd^2x\sqrt{g}\,\frac{1}{m}g^{ij}R^{-1}\partial_i R
	\partial_j(B+m\Omega)\,\,\,\,+o(\varepsilon^4)
	\,.
\end{equation}
An important point is that in the elliptic and hyperbolic cases, these actions are relevant at order $\varepsilon^2$, and hence must be included to the basis of independent invariants \cite{HS} used for constructing the most general effective action.  

Another useful observation regarding to what we are doing is as follows. It is straightforward to rewrite the formulas for the drift velocity 
and vorticity in terms of the minimal spin connection

\begin{equation}\label{drift-gr}
	\tilde{\cal V}^i=-2\,\e^{ij}\tilde{\cal E}_j/R\,,\qquad
	\tilde {\cal O}=\e^{ij}\partial_i\tilde{\cal V}_j\,.
\end{equation}
The idea behind of \eqref{drift-gr} is that $\tilde{\cal V}$ and $\tilde{\cal O}$ contains powers of $R^{-1}$. After combining these formulas with those for ${\cal V}$ and ${\cal O}$, we find that under nonrelativistic diffeomorphisms 

\begin{equation}\label{VO}
	\delta ({\cal O}-\tilde{\cal O})=-\partial_k({\cal O}-\tilde{\cal O})\xi^k
	\,,\qquad
	\delta({\cal V}^i-\tilde{\cal V}^i)=-\partial_k({\cal V}^i-\tilde{\cal V}^i)\xi^k+({\cal V}^k-\tilde{\cal V}^k)\partial_k\xi^i
\,.
\end{equation}
In other words, ${\cal O}-\tilde{\cal O}$ and ${\cal V}^i-\tilde{\cal V}^i$ are the scalar and vector, respectively.

Once we specify the new scalar field in terms of $\tilde{\cal O}$, we can construct invariant actions with higher powers of $R^{-1}$. The simplest 
example is $\int m({\cal O}-\tilde{\cal O})^2$. Its $1/R$ expansion reads

\begin{equation}\label{BO}
S=\int dtd^2x\sqrt{g}\,
m\bigl({\cal O}-\tilde{\cal O}\bigr)^2=\int dtd^2x\sqrt{g}\frac{1}{m}R^{-4}\Bigl[g^{ij}\partial_iR\partial_n B-R\,\triangle B\Bigr]^2
\,\,\,\,+o(\varepsilon^0)
\,,
\end{equation}
where $\triangle$ is the Laplace-Beltrami operator.

It is instructive to look at another invariant action

\begin{equation}\label{RO}
S=\int dtd^2x\sqrt{g}\,R
\bigl({\cal O}-\tilde{\cal O}\bigr)
=\int dtd^2x\sqrt{g}\biggl[2g^{ij}\frac{\partial_i R}{R}\tilde{\cal E}_j+R\Omega\biggr]\,\,\,\,+o(\varepsilon^4)
\,.
\end{equation}
After subtracting \eqref{R} and then using \eqref{BE}, we obtain 

\begin{equation}\label{invariant}
	S=\int dtd^2x\sqrt{g}\,R\Bigl[\frac{1}{m}B+\tilde{\cal O}\Bigr]
	=\int dtd^2x\sqrt{g}\,
	\biggl[\frac{1}{m}RB-2g^{ij}\frac{\partial_iR}{R}\tilde{\cal E}_j
\biggr]\,.
\end{equation}
This invariant action was derived in \cite{AHH} by manipulation of the variables, but the way we have presented above is much simpler and powerful. 

Though we have discussed the simple examples of the effective actions invariant under nonrelativistic diffeomorphisms, the discussion also makes clear that, like in general relativity, going to the most general form of the effective action requires knowledge of all objects such as scalars, vectors, tensors and their covariant derivatives. For example, one can consider a linear combination of several scalars ${\cal B}+aR+b\,m({\cal O}-\tilde{\cal O})$,  yielding a generalization of \eqref{R1} and \eqref{R2}. Certainly, this is an interesting direction to investigate in the future.

As mentioned above, on a Riemann surface one of the interesting questions is what the response of a physical system to a change in the curvature might 
be. A recent example that one might bear in mind is fractional quantum Hall states \cite{W} where, in particular, the leading terms in the gradient expansion of the particle density were obtained

\begin{equation}\label{current}
	J^0=J^0_0+\frac{1}{8\pi}R+\frac{{\text b}}{8\pi}B^{-1}\triangle R\,,\qquad {\text b}=\frac{7}{12}-\frac{1}{4\nu}
	\,,
\end{equation}
where $\nu$ is the filling fraction of Laughlin states. Note that $B$ is now a uniform magnetic field. 

Now the question arises: is the relative coefficient between the last two terms fixed by symmetry? The answer to this is clear from 
our analysis of the invariant actions. The first term comes from \eqref{WZ1} with $\kappa=1/2$, while the second from \eqref{R} and 
\eqref{RO}.\footnote{Both of those have $R\Omega$ which leads to $B^{-1}\triangle R$ in $J^0$.} These invariant actions are independent of 
each others, so they should be included into the basis of independent structures used for a construction of the most general effective action. 
Since, however, in this process the independent structures have arbitrary weights, the relative coefficient is also arbitrary. A possible way out 
is to impose more restrictions on the effective action. For example, one may require that it must be regular in the limit 
$m\rightarrow 0$ \cite{SonCo}. 

%________________________________________________________________________________________________________________________

\section{Turning spin on}
\renewcommand{\theequation}{5.\arabic{equation}}
\setcounter{equation}{0}
Now we would like to address some issues concerning effective actions for spinning particles (quasiparticles). 

In \cite{AHH}, we derived the Wards identities by considering a relativistic system in three dimensions, for which the energy-momentum tensor is conserved, and then taking a nonrelativistic limit of equation

\begin{equation}\label{T=0}
\nabla_\nu T^{\mu\nu}=0
\,.
\end{equation} 
Setting $\mu$ equal to $0$ and $i$ in this equation gives the current conservation law and continuity equation, respectively. However, in the case of spinning particles moving in a gravitational background the energy-momentum tensor is no longer conserved because of gravitational interaction. In the absence of torsion a gravitational force that acts on a spinning particle is given by $\tfrac{1}{2}R^{(3)}_{\mu\nu\lambda\sigma}\dot x^\nu S^{\lambda\sigma}$ \cite{van}, with $S^{\lambda\sigma}$ an antisymmetric $3$-tensor. It is called the intrinsic angular momentum (spin) tensor. As a consequence, equation \eqref{T=0} becomes 

\begin{equation}\label{T=R}
g_{\mu\nu}\nabla_\lambda T^{\lambda\nu}=\frac{1}{2}R^{(3)}_{\mu\nu\lambda\sigma}J^\nu S^{\lambda\sigma}
\,,
\end{equation}
where $J^{\mu}$ is the $4$-current.

Given equation \eqref{T=R}, it is straightforward to do a similar calculation and derive the corresponding equations generalizing those of \cite{AHH}. In doing so, we use the ansatz of \cite{AHH} for the $1/c$ expansion of the energy-momentum tensor

\begin{equation}\label{TJ}
	T^{\mu\nu}=
	\begin{pmatrix}
	\,\,mcJ^0\,\,&{}\,\,&mJ^i\,\,\\
	{}\,\,&{}\,\,&\,\,\\
	\,\,mJ^i\,\,&{}\,\,& T^{ij}/c\,\,
	\end{pmatrix}	
\,.
\end{equation}
In addition, we assume the following expansions for the current and the spin tensor

\begin{equation}\label{S}
	J^\mu=\Bigl(cJ^0,\,\,J^i\Bigr)
	\,,\qquad
	S^{\mu\nu}=\begin{pmatrix}
	\,\,0\,\,&{}\,\,&0\,\,\\
	{}\,\,&{}\,\,&\,\,\\
	\,\,0\,\,&{}\,\,& s\,\e^{ij}/c\,\,
	\end{pmatrix}
	\,,
	\end{equation}
where $s$ is a parameter (spin). 

First, we consider the time component of \eqref{T=R}. Then in the limit $c\rightarrow\infty$, \eqref{metricd+1}, \eqref{TJ}, and \eqref{S} give the current conservation law\footnote{Note that it can also be derived by taking a nonrelativistic limit of equation $\nabla_\mu J^\mu=0$.}

\begin{equation}\label{current-cons}
	\partial_tJ^0+\oh\partial_t\ln g J^0+\nabla_i J^i=0\,,
\end{equation}
as must be. This justifies us in assuming the $1/c$ expansion of $S^{\mu\nu}$.

Dealing with the spatial component will take a little more effort to obtain a final equation. Returning to \eqref{omega} we see that $2(\partial_k\omega_0-\dot\omega_k)=\e^{ij}\nabla_i\dot g_{kj}$. With this identity, we can proceed further. 
Using \eqref{metricd+1}, \eqref{TJ}, and \eqref{S}, we now arrive at 

\begin{equation}\label{stress-cons}
m\partial_tJ_i+\oh m\partial_t\ln gJ_i+\nabla_j T^j_i=-J^0\bigl(E_i+s\,\tilde{\cal E}_i\bigr)
+\text{e}_{ij}J^j\Bigl(B+\frac{s}{2}R\Bigr)
\,.
\end{equation}
An interesting observation one can make by a comparison with the $s=0$ case \cite{AHH} is that \eqref{stress-cons} is obtained by simply replacing: 
$A_0\rightarrow A'_0$ and $A_i\rightarrow A'_i$ with 

\begin{equation}\label{A'}
	A'_0=A_0+s\omega^0_0\,,\qquad
	A'_i=A_i+s\omega_i
	\,.
\end{equation}
It is worth noting that for any $s$ the $A'$s transform under nonrelativistic diffeomorphisms in the same way as the original gauge fields \eqref{transNR}. In \cite{AHH}, this fact was crucial for deriving the invariant actions. 

The assertion that an action is symmetric under nonrelativistic diffeomorphisms and $U(1)$ gauge transformations means that one can obtain the corresponding conservation laws by Noether's theorem. But now an opposite question arises as to how this can be done for \eqref{current-cons} and \eqref{stress-cons}? We can answer this as follows. We assume first that for a non-zero value of $s$ the effective action takes the form

\begin{equation}\label{Ss}
S(A_0,\,A_i,\,g_{ij},\,s)= S(A_0+s\omega^0_0,\,A_i+s\omega_i,\,g_{ij})
\,.
\end{equation}
In other words, $S$ is a functional of $A'_\mu$ and $g_{ij}$. The idea behind of this assumption is that such a form appears for actions invariant 
under the local $SO(2)\times U(1)$ symmetry group \cite{AHH}. Recently, it was also discussed in \cite{fradkin-new} regarding an inclusion of 
the spin structure in a covariant derivative. Now consider a variation of the action with respect to the background fields

\begin{equation}\label{variation}
	\delta S=\int dtd^2x\sqrt{g}
	\Bigl[J^0\bigl(\delta A_0+s\delta\omega^0_0\bigr)+J^i\bigl(\delta A_i+s\delta\omega_i\bigr)+\frac{1}{2}T^{ij}\delta g_{ij}
	\Bigr]
\,.
\end{equation}
The variation with respect to the spin connection is supposed to give a spin current. We assume that there is no more than one polarized particle (quasiparticle) species, so the spin current is proportional to the conserved $U(1)$ current. In this form, we can consider the variation with respect to the gauge transformations \eqref{transG}, getting the current conservation law \eqref{current-cons}. Repeating this for the $SO(2)$ gauge transformations \eqref{so2}, we find it again, as expected. One can analyze in a similar fashion the variation with respect to nonrelativistic diffeomorphisms \eqref{transNR}.\footnote{For the case $s=0$ this was done in \cite{AHH,son-moroz}.} So we get \eqref{stress-cons}, leading, after an identification $T^{0i}=mJ^i$, to the continuity equation 

\begin{equation}\label{continuity}
\partial_t T_i^0+\oh \partial_t\ln gT_i^0+\nabla_j T^j_i=\rho\Bigl(E_i+\frac{s}{2m}\partial_iB-\e_{ij}{\text v}^j\bigl(B+\frac{s}{2}R\bigr)+
s\bigl(\dot\omega_i-\partial_i\omega_0)
\Bigr)	
	\,.
\end{equation}
Here $J^\mu=(-\rho,-\rho {\text v}^i)$. The right hand side represents a superposition of gravitational and electromagnetic forces including the Lorentz force.

In conclusion, we will make a comment on invariant actions for $s\not=0$. A generalization of what we have done in sections II and IV is straightforward. 
As noted before, the replacement $A\rightarrow A'$ does the job. The reason for this is that such new gauge fields transform under nonrelativistic diffeomorphisms in the same way as the original ones. Moreover, the $U(1)$ gauge invariance of an effective action ensures invariance with respect 
to the local $SO(2)$ transformations.

%______________________________________________________________________________________________________________________________________________
\section{Concluding Comments}
\renewcommand{\theequation}{6.\arabic{equation}}
\setcounter{equation}{0}

(i) In some cases, going beyond the fourth order in $\varepsilon$ does not really change the discussion of section II, because the mapping \eqref{A} is valid up to $\varepsilon^4$. For example, if we consider the effective actions \eqref{R} and \eqref{dBdB} whose leading terms are of order $\varepsilon^2$, then we get 

\begin{equation}\label{R6}
S=
\int dtd^2x\sqrt{g}\,\frac{1}{m}R
\biggl[B+m\Omega+m^2\,\nabla^n\Bigl(\frac{\mathsf{d}v_n}{B}\Bigr)\biggr]
\,\,\,\,+o(\varepsilon^6)\,,
\end{equation}

\begin{equation}\label{dBdB6}
	\begin{split}
S=\int dtd^2x\sqrt{g}\,\frac{1}{m}g^{ij}B^{-1}
	\biggl[&\biggl(1-\frac{m}{B}\Omega+\frac{m^2}{B^2}\Omega^2-m^2\nabla^n\Bigl(\frac{\mathsf{d}v_n}{B}\Bigr)\biggr)
	\partial_iB\partial_jB \\
	&+2m\biggl(\partial_i\Omega+
	+m\nabla_i\nabla^n\Bigl(\frac{\mathsf{d}v_n}{B}\Bigr)
	-\frac{m}{2B}\partial_i\Omega^2
	\biggr)\partial_jB
	+m^2\partial_i\Omega\partial_j\Omega\biggr]\,\,\,\,+o(\varepsilon^6)
	\,,
\end{split}
\end{equation}
which are invariant up to order $\varepsilon^6$.	

(ii) There is more to say about the viscosity tensor if we generalize the analysis of Appendix C by considering the coordinate-dependent strain-rate tensor. 

In flat space our analysis carries over, with only a few changes. Just as before, the viscosity tensor takes the form \eqref{2order} and 
\eqref{2order'}, but now with the list of symmetric tensors to be completed by 

\begin{equation}\label{dd}
	t^6_{ij}=B^{-1}\delta_{ij}\partial^2
	\,,\qquad
	t^7_{ij}=B^{-1}\partial_i\partial_j
	\,,\qquad
	t^8_{ij}=mB^{-1}(v_i\partial_j+v_j\partial_i)
	\,,\qquad
	t^9_{ij}=B^{-2}(\partial_iB\partial_j+\partial_jB\partial_i)
	\,.
\end{equation}
Note that more detail concerning $t^6_{ij}$ and $t^7_{ij}$ can be found in \cite{AG}. 

In curved space the story is less simple. For an illustration, consider the invariant action \eqref{wdw}

\begin{equation}\label{wdw2} 
S=\frac{c}{2\pi}\int dtd^2x
	\biggl[\varepsilon^{\mu\nu\lambda}\omega_\mu\partial_\nu\omega_\lambda-\frac{1}{2m}\sqrt{g}RB
	\biggr]
\,.
\end{equation}
Like in section III, we have introduced the prefactor $c/2\pi$. 

Let a Riemann surface $\Sigma$ be elliptic or hyperbolic. We could choose the metric of constant curvature $\mathfrak{g}_{ij}(x)$ such that 
$\mathfrak{g}_{ij}=\mathfrak{e}^a_i\mathfrak{e}^a_j$ and study a linear response to a weak time dependent perturbation of the zweibein $e^a_i=\mathfrak{e}^a_i(x)+u^a_i(t)$. The metric now becomes $g_{ij}=\mathfrak{g}_{ij}+2u_{ij}+u_i^au_j^a$, with the coordinate dependent strain tensor $u_{ij}(t,x)=(\mathfrak{e}^a_iu^a_j+\mathfrak{e}^a_ju^a_i)/2$. Note that for $e^a_i=\delta^a_i$ these formulas reduce to those given in section III for the parabolic case. 

From this starting point, the analysis proceeds in an obvious way. The part of the action linear in $u\dot u$ is given by 

\begin{equation}\label{uu}
	S=\frac{c}{2\pi}\int dtd^2x\sqrt{\mathfrak{g}}\,\mathfrak{e}^{ij}u_{jn}
		\Bigl[\mathfrak{R}\,\mathfrak{g}^{nm}+2\,\mathfrak{e}^{nl}\mathfrak{e}^{km}\mathfrak{D}_l\mathfrak{D}_k
		\Bigr]\dot u_{im}
		\,,
\end{equation}
where we have dropped the terms involving a rotation tensor $M_{ij}=(\mathfrak{e}^a_iu^a_j-\mathfrak{e}^a_ju^a_i)/2$. We use fraktur letters to denote geometric objects associated with the unperturbed surface $\Sigma$ such as the metric $\mathfrak{g}_{ij}$, the scalar curvature $\mathfrak{R}$, the covariant derivative $\mathfrak{D}_i$, and so on. Given this, the stress tensor response, linear in $\dot u$, can be read from the result of variation $\delta S/\delta u_{ij}$. We find

\begin{equation}\label{Tij1}
	T^{ij}=-\frac{c}{\pi}
	\Bigl[
	\bigl(\mathfrak{e}^{im}\mathfrak{g}^{jn}+\mathfrak{e}^{jm}\mathfrak{g}^{in}\bigr)
	\bigl(\mathfrak{R}-\mathfrak{D}_k\mathfrak{D}^k\bigr)+\mathfrak{e}^{im}\mathfrak{D}^{(j}\mathfrak{D}^{n)}+
	\mathfrak{e}^{jm}\mathfrak{D}^{(i}\mathfrak{D}^{n)}
	\Bigr]\dot u_{nm}
\,.
\end{equation}
With such a definition of $T^{ij}$, we need to take into account a relative minus sign between \eqref{Tij1} and \eqref{viscosity}. This results from the formula $\delta S=\frac{1}{2}\int dtd^2x\sqrt{g}\,T^{ij}\delta g_{ij}=\frac{1}{2}\int dtd^2x\sqrt{g}\,T_{ij}\delta g^{ij}$, with $\delta g^{ij}=-g^{in}g^{jm}\delta g_{nm}$. Therefore $T^{ij}$ takes the form

\begin{equation}\label{Tij}
	T^{ij}=-\eta^{ijnm}\dot u_{nm}+\dots
	\,\,,
\end{equation}
where $\eta^{ijmn}$ is the viscosity tensor. 

As usual, the viscosity tensor is divided into a symmetric and antisymmetric part under exchange of the first and the last pair of indices. 
To order $\varepsilon^2$, a general structure of $\eta_A$ can be analyzed along the lines of Appendix C, but we will not do it here. 
Instead, we only present what is relevant for the problem at hand. It is just that

\begin{equation}\label{etaA}
	\eta^{ijnm}_A=\frac{1}{2}\sum_{I=0,6,7}
	\eta_I^{\text{\tiny H}}\bigl(\mathfrak{e}^{im}t^{jn}_I+\mathfrak{e}^{jn}t^{im}_I+
	\mathfrak{e}^{jm}t^{in}_I+\mathfrak{e}^{in}t^{jm}_I\bigr)+\dots
\,,
\end{equation}
with
\begin{equation}\label{etaA1}
	t^{ij}_0=\mathfrak{g}^{ij}\,,\qquad
	t^{ij}_6=\mathfrak{R}^{-1}\mathfrak{g}^{ij}\mathfrak{D}_k\mathfrak{D}^k
	\,,\qquad
	t^{ij}_7=\mathfrak{R}^{-1}\mathfrak{D}^{(i}\mathfrak{D}^{j)}
	\,.
\end{equation}
Note that for $\mathfrak{g}_{ij}=\delta_{ij}$ the differential operators in these symmetric tensors reduce to those of \eqref{dd}. This is the reason why we enumerate them in the same way. The lack of the Ricci tensor in \eqref{etaA1} is due to the relation $\mathfrak{R}_{ij}=\tfrac{1}{2}\mathfrak{R}\mathfrak{g}_{ij}$ valid in two dimensions.

It follows now by comparison with \eqref{Tij1} that 

\begin{equation}\label{etaA2}
	\eta^{\text{\tiny H}}(c)=-\eta_6^{\text{\tiny H}}(c)=\eta_7^{\text{\tiny H}}(c)=
	\frac{c}{\pi}\,{\mathfrak R}
	\,.
\end{equation}

(iii) Effective actions can be computed by integrating out dynamical degrees of freedom. In the path integral formulation it is schematically given by 

\begin{equation}\label{integral} 
\e^{iS(A,g,m)}=\int D\Psi\,\e^{iS_0(\Psi,A_0,g_0,m_0)}
\,. 
\end{equation}
For the sake of simplicity, we use the same notation as in the case above, but now $\{A,g,m\}$ and $\{A_0,g_0,m_0\}$ describe the renormalized and bare parameters, respectively. 

If the theory under consideration is free from divergences, one has no need to use any regularization. In this case, the renormalized parameters simply coincide with the bare ones that allows one to equally discuss the underlying symmetries on both the sides of \eqref{integral}. Several original examples explored in the literature \cite{SW,SonCo} suggest this for the minimal set of the background fields $g_{ij}$ and $A_\mu$. 

In practice, it is more often that divergences appear. Let us assume that the underlying theory is renormalizable. Here one can use different renormalization schemes. With one scheme, the effective action may have the nonrelativistic symmetry generated by \eqref{transNR}. With another, 
it may have a more complicated symmetry structure like that of \cite{SonCo}. The point is that effective actions computed by different schemes are related by coupling constant redefinition. It is important to keep that in mind when comparing two different effective actions of the same quantum field theory.

So far, we have not made any assumption about the underlying theory. What would happen if the underlying theory is divergent? Although this is not the aim of our study, a couple of points is worthy of note.

First, for infinitesimal transformations \eqref{transNR} the limit $m\rightarrow 0$ is well-defined. However, if $m$ is the only parameter of dimension of mass, the problem may appear in specific effective actions requiring dimensionful couplings. For example, this happens in \eqref{BB}, where $S=
\int{\cal B}^2/m$. If there exists another parameter of dimension of mass, then the problem may be gone. In this case \eqref{BB} becomes $S=\int{\cal B}^2/\mu$, with $\mu$ a dynamically generated scale in \eqref{integral}. Of course, one can construct more examples of invariant effective actions along the lines of section IV by considering $A=\lim_{m\rightarrow 0}{\cal A}$. Note that in the limit $m\rightarrow 0$ all the effective actions discussed above become ill-defined or trivial, with only one exception. It is the extension \eqref{WZ} of the Wen-Zee action. This is an indication that its relativistic counterpart also provides the viscosity-conductivity relation of \cite{HS,Read}.

Second, the requirement that an effective action is invariant under nonrelativistic diffeomorphisms does not unambiguously fix a renormalization scheme. A field redefinition of the form \eqref{A'} respects the symmetry and is an example of a transformation of coupling parameters which relates effective actions computed in two different renormalization schemes. Let us see how this works for the effective action of \cite{HS}. It is that 

\begin{equation}\label{action}
	S(A,g)=\sum_{i=1}^5 a_i S_i(A,g)+o(\varepsilon^2)
	\,,
\end{equation}
where the $S_i$'s are defined in section II. Alternatively, the action can be written in terms of $A'$ with the 
corresponding coefficients $a'$. The field redefinition \eqref{A'}, with $s$ an arbitrary parameter, relates them such that 

\begin{equation}\label{transformations}
	a_1=a_1'\,,\quad
	a_2=-s a_1'+a_2'\,,\quad
	a_3=2s a_1'+a_3'\,,\quad
	a_4=-\frac{1}{2}s^2a_1'+s a_2'-\frac{1}{4}s a_3'+a_4'\,,\quad
	a_5=\frac{1}{4}s^2 a_1'-s a_2'-\frac{1}{4}s a_3'+a_5'\,.
\end{equation}
From \eqref{transformations} we find that, in addition to $a_1$, a linear combination 

\begin{equation}\label{inv}
2a_2+a_3=2a_2'+a_3'
\end{equation}
is scheme-independent.

One can analyze in a similar fashion a field redefinition 

\begin{equation}\label{A"}
%	\begin{split}
	A'_0=A_0+\mu{\cal V}^i(\tilde{\cal V}_i-{\cal V}_i)=
	A_0-\mu\Bigl(v_iv^i+\frac{2}{R}\e^{ij}v_i\tilde{\cal E}_j
	%\Bigl(\dot\omega_j-\partial_j\omega_0+\frac{1}{2m}\partial_jB\Bigr)
	\Bigr)+o(\varepsilon^2)
	\,,\quad
	A'_i=A_i+\mu({\cal V}_i-\tilde{\cal V}_i)=
	A_i+\mu\Bigl(
	v_i+\frac{2}{R}J_i^j\tilde{\cal E}_j
	%\Bigl(\dot\omega_j-\partial_j\omega_0+\frac{1}{2m}\partial_jB\Bigr)
	\Bigr)+o(\varepsilon)
	\,,
%\end{split}
\end{equation}
where $\mu$ is a parameter of mass dimension. We will not do so here. However, it is worth noting that this redefinition results in the following relation 
between the coefficients in front of the $R\Omega$ term in \eqref{R}: $a_4=(1+\mu/m)a_4'$. This is an indication on the scheme dependence of ${\text b}$ in \eqref{current}.

(iv) Another idea for the transformation rules that may be accessible by the tools developed in this work is that of \cite{SonCo, NC}

\begin{equation}\label{transNR2}
\delta A_0=-\partial_k A_0\xi^k- A_k\dot\xi^k+\frac{1}{4}(2s-g)J^i_k\nabla_i\dot\xi^k
\,,\quad
\delta A_i=-\partial_k A_i\xi^k- A_k\partial_i\xi^k-mg_{ik}\dot\xi^k\,,\quad
\delta g_{ij}=-\partial_k g_{ij}\xi^k-g_{kj}\partial_i\xi^k-g_{ik}\partial_j\xi^k\,,
\end{equation}
where $s$ and $g$ are some parameters. In fact, there is no need to redo all the calculations again. One can achieve the desired result just by making the following field redefinition \cite{SonCo, AHH}

\begin{equation}
A_0\rightarrow A_0+\frac{g}{4m}B-s\omega_0\,,\qquad
A_i\rightarrow A_i-s\omega_i
\,.
\end{equation}

(v) Recently, the minimal set of the background fields has been extended to describe sources for the energy density and current \cite{NC}. There has also 
been a discussion of the Newton-Cartan geometry with torsion and other relevant discussions of Ward identities, thermal transport, and 
Galilei-invariant systems \cite{Sfol}.

%__________________________________________________________________
\begin{acknowledgments}

We would like to thank M. Haack and S. Hofmann for discussions and collaboration during the initial stages of this work. We also wish to thank 
I.S. Burmistrov and G.E. Volovik for helpful discussions. Finally, we would like to thank the Arnold 
Sommerfeld Center for Theoretical Physics at Ludwig Maximilians Universit\"at for the warm hospitality. This research was supported in part by the DFG grant HA 3448/3-1 and the Alexander von Humboldt Foundation.

 \end{acknowledgments}

%__________________________________________________________________
\appendix
\section{Notation and Useful Formulas}
\label{notation}
\renewcommand{\theequation}{A.\arabic{equation}}
\setcounter{equation}{0}

Throughout this paper we use the same notation as in \cite{AHH}. In addition, indices $A$, $B$, etc. label coordinates in a locally inertial 
(tangent) coordinate system. Thus, $\eta_{AB}$ is the usual Minkowski metric, of signature $(-,+,\dots,+)$. In three dimensions, $\varepsilon^{ABC}$ is a totally antisymmetric tensor normalized by $\varepsilon^{012}=1$. 

Up to $1/c^3$ terms, the $1/c$ expansions of the fields \eqref{bu} are given by 

\begin{equation}\label{bu1}
	b={\cal B}\biggl(1-\frac{1}{2(mc)^2}{\cal A}_i{\cal A}^i
	\biggr)
	\,,\qquad
	u_0=-1+\frac{1}{mc^2}{\cal A}_0
	\,,
	\qquad
	u_i=\frac{1}{mc}{\cal A}_i
	\,.
\end{equation}
Here we have used the relation \eqref{A-}. Note that these expansions are not gauge invariant that may seem counterintuitive. The reason for that is 
that the $1/c$ expansion of the metric \eqref{metricd+1} is not gauge invariant.

In the case of the dreibein, the $1/c$ expansions are 

\begin{equation}\label{e}
\e_\mu^A=
\begin{pmatrix}
1-{\displaystyle\frac{A_0}{mc^2}}+{\displaystyle\frac{1}{2}\frac{A_iA^i}{(mc)^2}}\,\,&{}\,\,&{\displaystyle\frac{1}{mc}A^k e^a_k}\,\,\\
{}\,\,&{}\,\,&\,\,\\
{\displaystyle 0}\,\,&{}\,\,& e^a_i\,\,
\end{pmatrix}\,,
\qquad\qquad
\e^\mu_A=
\begin{pmatrix}
1+{\displaystyle\frac{A_0}{mc^2}}-{\displaystyle\frac{1}{2}\frac{A_iA^i}{(mc)^2}}\,\,&{}\,\,&{-\displaystyle\frac{1}{mc}A^i}\,\,\\
{}\,\,&{}\,\,&\,\,\\
{\displaystyle 0}\,\,&{}\,\,& e_a^i\,\,
\end{pmatrix}\,,
\end{equation}
where $e^a_i$ is a zweibein such that $g_{ij}=e^a_i e^a_j$. The corresponding expansion of the minimal spin connection then takes the form 

\begin{equation}\label{spin-con}
	{\text w}_\mu^{0a}=0\,,
	\qquad
	{\text w}_0^{ab}=\frac{1}{c}\varepsilon^{ab}\omega^0_0=
	\frac{1}{c}\varepsilon^{ab}\Bigl(\omega_0-\frac{1}{2m}B\Bigr)\,,
	\qquad
	{\text w}_i^{ab}=\varepsilon^{ab}\omega_i\,,
\end{equation}
with $\omega_0$ and $\omega_i$ given by \eqref{omega}. The $\omega$'s satisfy
	
\begin{equation}\label{dR}
	2\e^{ij}\nabla_i(\dot\omega_j-\partial_j\omega_0)=\dot R+\frac{1}{2}\partial_t\ln g R
	\,,
\end{equation}
where $R$ is the scalar curvature and $g=\det g_{ij}$.

The fields $\omega^0_0$ and $\omega_i$ transform under nonrelativistic diffeomorphisms similar to the gauge fields ${\cal A}_0$ and ${\cal A}_i$

	\begin{equation}\label{spin-trans}
		\delta\omega^0_0=-\partial_k\omega^0_0\xi^k-\omega_k\dot\xi^k\,,\qquad
		\delta\omega_i=-\partial_k\omega_i\xi^k-\omega_k\partial_i\xi^k\,.
	\end{equation}
In addition, under the $SO(2)$ rotations $\delta e^a_i=\Lambda\varepsilon^{ab}e^b_i$, the transformation rules are 

\begin{equation}\label{so2}
	\delta\omega^0_0=-\dot\Lambda\,,\qquad
	\delta\omega_i=-\partial_i\Lambda\,.
\end{equation}

For the drift velocity and vorticity defined in terms of the minimal spin connection \eqref{drift-gr}, the transformation rules are given by 
\begin{equation}\label{drift-gr2}
	\delta\tilde{\cal V}^i=-\partial_k\tilde{\cal V}^i\xi^k+\tilde{\cal V}^k\partial_k\xi^i+\dot\xi^i\,,\quad
	\delta\tilde {\cal O}=-\partial_k\tilde {\cal O}\xi^k-J^i_k\nabla_i\dot\xi^k\,.
\end{equation}

In addition to \eqref{A}, it is also useful to have the following formulas at hand 

\begin{equation}\label{B}
	{\cal B}=B+m\Omega+m^2\,\nabla^n\Bigl(\frac{\mathsf{d}v_n}{B}\Bigr)+O(\varepsilon^6)
	\,,\qquad
	{\cal E}_i=E_i+m\bigl(\dot v_i+v^k\nabla_i v_k
	\bigr)+O(\varepsilon^5)
	\,\,
\end{equation}
together with those for the inverse mappings

\begin{equation}\label{BE}
B={\cal B}-m{\cal O}
\,,
\qquad
E_i={\cal E}_i-m\bigl(\dot{\cal V}_i+{\cal V}^k\nabla_i{\cal V}_k\bigr)
\,.
\end{equation}

%__________________________________________________________________
\section{An Example of ${\cal A}(A)$ Mapping}
\renewcommand{\theequation}{B.\arabic{equation}}
\setcounter{equation}{0}

Here we give an example of solving equations \eqref{A-m}. This will enable us to show how a field redefinition works in a concrete example.

In fact, if ${\cal Q}_i$ vanishes, then ${\cal Q}_0$ transforms as a scalar under nonrelativistic diffeomorphisms. A typical example of the scalar field that immediately comes to mind is ${\cal B}$.\footnote{Another option is to use ${\cal O}-\tilde{\cal O}$, as it follows from \eqref{VO}. In this case, \eqref{A-m} becomes $A_0={\cal A}_0+\tfrac{1}{2}m{\cal V}_i{\cal V}^i+\alpha({\cal O}-\tilde{\cal O})$ and $A_i={\cal A}_i-m{\cal V}_i$.} Given this, equations \eqref{A-m} take the form

\begin{equation}\label{A-ex}
A_0={\cal A}_0+\frac{1}{2}m{\cal V}_i{\cal V}^i+\frac{\alpha}{m}{\cal B}
	\,,\qquad
	A_i={\cal A}_i-m{\cal V}_i
	\,,
\end{equation}
where $\alpha$ is a numerical factor, introduced for convenience. Within the $\varepsilon$-expansion, these equations are solved by 

\begin{equation}\label{A2}
{\cal A}_0=A_0-\alpha\Bigl(\frac{B}{m}+\Omega\Bigr)-
\frac{m}{2B^2}g^{ij}E'_iE'_j-
\frac{\alpha^2}{m}\nabla^i\Bigl(\frac{\partial_i B}{B}\Bigr)+O(\varepsilon^4)
\,,\qquad
{\cal A}_i=A_i-\frac{m}{B}J^k_iE'_k+O(\varepsilon^3)
\,,
\end{equation}
where $E'_i=E_i+\tfrac{\alpha}{m}\partial_iB$.

Now we want to see how this solution is related to that given by \eqref{A}. In doing so, we make the field redefinition 

\begin{equation}\label{fr}
A_0\rightarrow A_0-\alpha\Bigl(\frac{B}{m}+\Omega\Bigr)-\frac{\alpha^2}{m}\nabla^i\Bigl(\frac{\partial_i B}{B}\Bigr)\,,\qquad
A_i\rightarrow A_i
\end{equation}
in \eqref{A} that does lead to \eqref{A2}, to the given order in $\varepsilon$.

In general, equations of motion may be used in proving the invariance of a field (operator) under a symmetry. This is the case for the example we have been considering. Indeed, the equations of motion for ${\cal A}_\mu$ derived from the following action invariant under nonrelativistic diffeomorphisms 

\begin{equation}\label{axac}
S=\int dtd^2x\sqrt{g}
\biggl[\frac{\varepsilon^{\mu\nu\lambda}}{\sqrt{g}}\Bigl({\cal A}_\mu-2A_\mu\Bigr)\partial_\nu{\cal A}_\lambda-\frac{m}{{\cal B}}g^{ij}{\cal E}_i{\cal E}_j+\frac{\alpha}{m}{\cal B}^2\biggr]
\end{equation}
are
\begin{equation}\label{axac2}
B={\cal B}-m{\cal O}
\,,
\qquad
E_i={\cal E}_i-m\bigl(\dot{\cal V}_i+{\cal V}^k\nabla_i{\cal V}_k\bigr)-\frac{\alpha}{m}\partial_i{\cal B}
\,.
\end{equation}
These are equivalent to equations \eqref{A-ex}, modulo gauge transformation.

Finally, let us note that two different mappings have been recently discussed in \cite{wuwu}. However, the issue of field redefinition was not raised.

%__________________________________________________________________
\section{Viscosity Tensor and $\varepsilon$-Expansion}
\label{v-tensor}
\renewcommand{\theequation}{C.\arabic{equation}}
\setcounter{equation}{0}

The viscosity tensor $\eta$ is a tensor of rank $4$ which relates the stress tensor $T$ to the strain-rate tensor $\dot u$. In the limit of small strain-rates, the relation is linear

\begin{equation}\label{viscosity}
	T_{ij}=\eta_{ijnm}\dot u_{nm}+\dots
	\,\,.
\end{equation}
Here the dots mean higher derivative terms, and the geometry is Euclidean. It is useful to divide $\eta$ into a symmetric and antisymmetric part under exchange of the first and the last pair of indices \cite{zograf}

\begin{equation}\label{split}
	\eta_{ijnm}=\ets_{ijnm}+\etaa_{ijnm}\,,\quad
\ets_{ijnm}=\ets_{nmij}\,,\quad
\etaa_{ijnm}=-\etaa_{nmij}
	\,.
\end{equation}

In two-dimensional flat space, if the tensors $T$ and $\dot u$ are symmetric and rotational symmetry is not broken, the viscosity tensor has only three independent components. Explicitly, it is given by 

\begin{equation}\label{0order}
\ets_{ijnm}=\bigl(\zeta-\eta^{\text{\tiny S}}\bigr)
\delta_{ij}\delta_{nm}+
\ets
\bigl(\delta_{in}\delta_{jm}+\delta_{im}\delta_{jn}\bigr)
\,,\quad	
\etaa_{ijnm}=\frac{1}{2}\eta^{\text{\tiny H}}\bigl(\delta_{jn}\varepsilon_{im}+\delta_{im}\varepsilon_{jn}+
\delta_{in}\varepsilon_{jm}+\delta_{jm}\varepsilon_{in}\bigr)
\,,
\end{equation}
with $\zeta$ the bulk viscosity, $\ets$ the shear viscosity, and $\eta^{\text{\tiny H}}$ the Hall viscosity.

In our present discussion, we consider invariant actions within the $\varepsilon$-expansion of \cite{HS}. It is clear that this type of 
expansion would arise for the stress tensor as long as one uses $T_{ij}=\tfrac{2}{\sqrt{g}}\delta S/\delta g^{ij}$. 

What would the resulting expansion of the viscosity tensor look like? If an action is of order $\varepsilon^2$, then the viscosity tensor is of 
order zero. In this case, it is given by \eqref{0order}, where all the coefficients are of order $\varepsilon^0$. If one has an action up to order $\varepsilon^4$, the viscosity tensor will be of order $\varepsilon^2$. For $\dot u$ independent of coordinates\footnote{Or $\eta_{ijnm}$ is not a differential operator including spatial derivatives.}, a simple analysis shows that in flat space the viscosity tensor takes the form 

\begin{align}\label{2order}
\ets_{ijnm}=&\frac{1}{2}\sum_{I=0}^5 
\bigl(\zeta_I-\ets_I\bigr)\bigl(\delta_{ij}t_{nm}^I+\delta_{nm}t_{ij}^I\bigr)+
\ets_I
\bigl(\delta_{im}t^I_{jn}+\delta_{jm}t^I_{in}
+\delta_{in}t^I_{jm}+\delta_{jn}t^I_{im}
\bigr)
\,,\quad	\\
\label{2order'}
\etaa_{ijnm}=&\frac{1}{2}\sum_{I=0}^5 \eta^{\text{\tiny V}}_I\bigl(\delta_{ij}t^I_{nm}-\delta_{nm}t^I_{ij} \bigr)+
\eta^{\text{\tiny H}}_I
\bigl(\varepsilon_{im}t_{jn}^I+\varepsilon_{jn}t_{im}^I+
\varepsilon_{jm}t_{in}^I+\varepsilon_{in}t_{jm}^I \bigr)
\,,
\end{align}
with 
\begin{align*}
    \qquad\qquad\qquad\qquad\qquad
    t_{ij}^0&=\delta_{ij}\,&,\qquad
	t_{ij}^1&=m\left(\partial_iBv_j+\partial_jBv_i\right)/B^2\,&,\qquad
	t_{ij}^2&=m^2v_iv_j/B
	\,&,\qquad\qquad\qquad\qquad\qquad\qquad\qquad\qquad\qquad
	\\
	\qquad\qquad\qquad\qquad\qquad
	t_{ij}^3&=\partial_i\partial_j B/B^2\,&,\qquad
	t_{ij}^4&=m\left(\partial_iv_j+\partial_jv_i\right)/B\,&,\qquad
	t_{ij}^5&=\partial_iB\partial_jB/B^3
	\,&.\qquad\qquad\qquad\qquad\qquad\qquad\qquad\qquad\qquad
\end{align*}
Here we have introduced a set of symmetric tensors and normalized them to be dimensionless. Thus, the viscosity tensor now contains $23$ independent parameters.\footnote{Note that $\eta^{\text{\tiny V}}_0\equiv 0$.} The parameters labeled by $I=1,\dots,5$ are of order $\varepsilon^0$. The remaining parameters may be of order $\varepsilon^2$. Obviously, 
they represent the zero order bulk, shear, and Hall viscosities \eqref{0order} accompanied by $\varepsilon^2$-corrections. Therefore we set $\zeta_0=\zeta$, $\ets_0=\ets$, and $\eta^{\text{\tiny H}}_0=\eta^{\text{\tiny H}}$.
%___________________________________________________________________

\end{document}